\documentclass{elsart}

 \usepackage{graphicx}

\usepackage{amssymb}

\begin{document}

\begin{frontmatter}



\title{Phase transition of an extrinsic curvature model on tori}


\author[label1]{Hiroshi Koibuchi}
\ead{koibuchi@mech.ibaraki-ct.ac.jp}

\address[label1]{Department of Mechanical and Systems Engineering, Ibaraki National College of Technology, 
Nakane 866, Hitachinaka,  Ibaraki 312-8508, Japan}

\begin{abstract}
We show a numerical evidence that a tethered surface model with extrinsic curvature undergoes a first-order crumpling transition between the smooth phase and a non-smooth phase on triangulated tori. The results obtained in this Letter together with the previous ones on spherical surfaces lead us to conclude that the tethered surface model undergoes a first-order transition on compact surfaces.
\end{abstract}

\end{frontmatter}

\section{Introduction}\label{intro}
The tethered surface model of Helfrich, Polyakov, and Kleinert (HPK) \cite{HELFRICH-NF-1973,POLYAKOV-NPB-1986,Kleinert-PLB-1986} is a surface model for artificial and biological membranes and for strings \cite{NELSON-SMMS-2004,Gompper-Schick-PTC-1994,BOWICK-TRAVESSET-PREP-2001,WIESE-PTCP19-2000,WHEATER-JP-1994}. The vertices, which represent the lipid molecules of membranes, are prohibited from the free diffusion over the surface because of the fixed connectivity nature of the triangulated surface. The phase transition between the smooth phase and a non-smooth phase (or crumpled phase) of the model  has long been studied, and the phase structure is being clarified step by step \cite{KOIB-PLA-2003-2,KOIB-EPJB-2004,KOIB-PLA-2005-1,KOIB-PLA-2006-1,BCFTA-1996-1997,WHEATER-NPB-1996,KANTOR-NELSON-PRA-1987,GOMPPER-KROLL-PRE-1995,KOIB-PRE-2003,KOIB-PRE-2004-2,Kownacki-Diep-PRE-2002,KOIB-PRE-2004-1,KOIB-PRE-2005,KOIB-NPB-2006,KANTOR-SMMS-2004,KANTOR-KARDER-NELSON-PRA-1987,BCTT-EPJE-2001,BOWICK-SMMS-2004,NELSON-SMMS-2004-2}.

An intrinsic curvature model for membranes is known to undergo a first-order crumpling transition, which was confirmed by Monte Carlo (MC) simulations \cite{KOIB-EPJB-2004,KOIB-PLA-2005-1,KOIB-PLA-2006-1}. The transition occurs on a disk  \cite{KOIB-PLA-2005-1}, which is a noncompact surface. Moreover, the transition can be seen on a sphere \cite{KOIB-EPJB-2004} as well as on a torus \cite{KOIB-PLA-2006-1}. Therefore, we can state that the tethered surface model with intrinsic curvature has a first-order transition on an arbitrary surface in ${\bf R}^3$. 

However, it remains to be stated that the extrinsic curvature model has a first-order transition on compact surfaces. In fact, the first-order transition of the model has never been confirmed on a torus, although it has been confirmed on a sphere by the MC simulations \cite{KOIB-PRE-2004-1,KOIB-PRE-2005,KOIB-NPB-2006}. The reason why it is so hard to clarify the order of the transition on a torus seems because of the size effect. The extrinsic curvature model shows the first-order transition only on the sphere of size $N\!\geq\! 7000$ \cite{KOIB-PRE-2005,KOIB-NPB-2006}, which is very large in contract to that in the model with intrinsic curvature, where the transition can be seen even on very small surfaces \cite{KOIB-EPJB-2004,KOIB-PLA-2005-1,KOIB-PLA-2006-1}. 

\begin{figure}[htbp]
\centering
(a)\includegraphics[width=50mm]{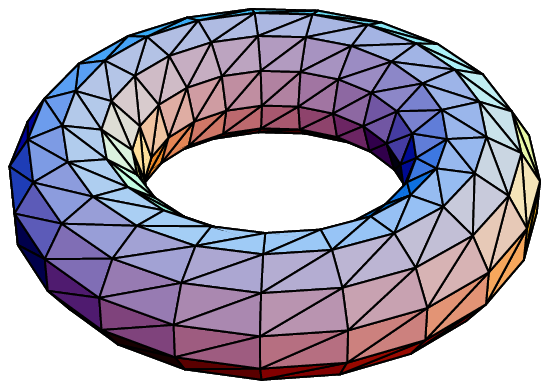}
(b)\includegraphics[width=50mm]{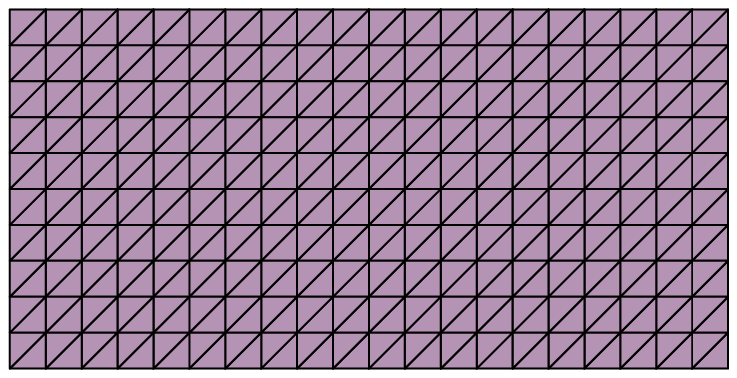}
  \caption{(a) The real torus obtained by connecting two parallel sides of (b) the rectangular surface of size $L_1\!\times\!L_2\!=\!20\!\times\!10$.}
  \label{fig-1}
\end{figure}

It is true that the size effect disappears at sufficiently small curvatures on the sphere because the first-order transition occurs only at large $N$, which corresponds to the large radius $R$ of the sphere. Because the principal curvature is given by $\kappa\!=\!1/R$, we have $\kappa=\sqrt{4\pi \over N}$, where the area of the surface is identified with $N$. On the torus in Fig. \ref{fig-1}(a) made from the square of size $(L_1,L_2)$ shown in Fig. \ref{fig-1}(b) we have  approximately $\kappa=\sqrt{8\pi^2 \over N}$, which is the larger one of the two principal curvatures, where $L_1\!=\!2L_2$ and $2\pi R_i\!=\!L_i (i\!=\!1,2)$ were assumed. Therefore, the first-order transition can be seen on the torus of size $6\!\sim\!7$ times larger than that of the sphere, because the sphere and the torus are considered to have the common $\kappa$ where the size effect disappears. It is natural to consider that low-frequency modes of surface fluctuations play an important role in the crumpling transition, and that the corresponding energy scales are independent of the surface topology.

In this Letter, we study the extrinsic curvature model in \cite{KOIB-PRE-2004-1}, where the first-order transition was seen even on the sphere of size $N=1500\!\sim\!2500$, which are relatively smaller than $N\!\simeq\! 7000$ where the first-order transition was observed in the model with the standard bending energy. The bending energy in \cite{KOIB-PRE-2004-1} is slightly different from the standard one, however, the model is in the same class as that of the standard bending energy.  

We will show in this Letter the first numerical evidence that a tethered surface model with extrinsic curvature undergoes a first-order crumpling transition between the smooth phase and a non-smooth phase on triangulated surfaces of torus topology. It was already reported by us that the tethered surface model of HPK undergoes a first-order transition on spherical surfaces \cite{KOIB-PRE-2004-1,KOIB-PRE-2005,KOIB-NPB-2006}. The results obtained in this Letter together with the previous ones \cite{KOIB-PRE-2004-1,KOIB-PRE-2005,KOIB-NPB-2006} lead us to conclude that the tethered surface model of HPK undergoes a first-order crumpling transition on {\it compact} surfaces.

\section{Model}\label{model}
In order to define the bending energy $S_2$, we use the normal vector of the vertex $i$ such as
\begin{equation}
\label{normal}
{\bf n}(i)= {{\bf N}_i \over \vert {\bf N}_i\vert}, \quad {\bf N}_i = \sum _{j(i)} {\bf  n}_{j(i)} A_{\it \Delta_{j(i)}},
\end{equation}
where $\sum _{j(i)}$ denotes the summation over triangles  $j(i)$ linked to the vertex $i$. The vector ${\bf n}_{j(i)}$ is the unit normal of the triangle $j(i)$, and  $A_{\it \Delta_{j(i)}}$ is the area of $j(i)$.

The Gaussian tethering potential $S_1$ and the extrinsic curvature $S_2$ are defined by
\begin{equation}
\label{S1S2}
S_1=\sum_{(ij)} \left(X_i-X_j\right)^2, \quad
S_2=\sum_i\sum_{j(i)}\left[1-{\bf n}(i)\cdot {\bf n}_{j(i)}\right],
\end{equation}
where $\sum_{(ij)}$ is the sum over bond $(ij)$ connecting the vertices $X_i$ and $X_j$.  It should be noted that $S_2({\rm illdef})=\sum_{i,j}\left(1-{\bf n}(i)\cdot {\bf n}(j)\right)$ defined only by using the normal vectors ${\bf n}(i)$ in Eq.(\ref{normal}) is not well defined. This ill-definedness comes from the fact that there exist two surfaces locally different from each other that has the same value of $S_2({\rm illdef})$. Two normal vectors at the ends of a bond $(i,j)$ can be parallel for surfaces that are not smooth.

\begin{figure}[hbt]
\centering
\includegraphics[width=5cm]{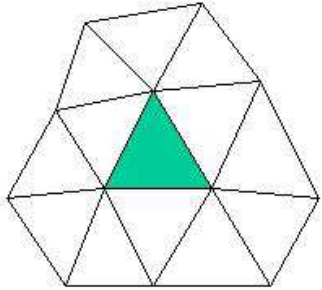}
\caption{The range of the interaction between the normal vectors in $S_2$. The normal of the shaded triangle interacts with the normals of the triangles shown in the figure.}
\label{fig-2}
\end{figure}

In order to understand graphically the interaction described in Eq. (\ref{S1S2}), we show the interaction range between the normals of triangles in Fig. \ref{fig-2}. The normal of the shaded triangle interacts with the normals of the triangles in the figure.

The partition function is defined by 
\begin{eqnarray}
 \label{partition-function}
Z(b) = \int \prod _{i=1}^N dX_i \exp\left[-S(X)\right],\qquad \\
S(X)=S_1 + b S_2,  \qquad\quad \nonumber
\end{eqnarray}
where $N$ is the total number of vertices as described above. The expression $S(X)$  shows that $S$ explicitly depends on the variable $X$. The coefficient $b$ is the bending rigidity. The surfaces are allowed to self-intersect. The center of surface is fixed in $Z(b) $ to remove the translational zero-mode.

\section{Monte Carlo technique}\label{MC-Techniques}
The canonical Monte Carlo technique is used to update the variables $X$ so that $X^\prime \!=\! X \!+\! \delta X$, where the small change $\delta X$ is made at random in a small sphere in ${\bf R}^3$. The radius $\delta r$ of the small sphere is chosen at the beginning of the simulations to maintain the rate of acceptance $r_X$ for the $X$-update as $r_X \simeq 0.5$. 

We use a random number called Mersenne Twister \cite{Matsumoto-Nishimura-1998} in the MC simulations. Single sequence of random number is used for 3-dimensional move of vertices $X$ and  for the Metropolis accept/reject in the update of $X$.

 We use surfaces of size $N\!=\!1800$, $N\!=\!3200$, $N\!=\!5000$,  $N\!=\!9800$ and  $N\!=\!16200$ for $L_1/L_2\!=\!2$,  and $N\!=\!1764$, $N\!=\!3600$, $N\!=\!6400$,  $N\!=\!10000$ and  $N\!=\!16900$ for $L_1/L_2\!=\!4$.  The size of the surfaces is obviously larger than that of the same model on spherical surfaces used in \cite{KOIB-PRE-2004-1}.

The total number of MCS after the thermalization MCS is about $2\!\times\!10^8\sim 2.6\!\times\!10^8$ at the transition point of surfaces of size $N\!=\!9800\sim N\!=\!16900$,  and $1.0\!\times\!10^8\sim 1.6\!\times\!10^8$ for  $N\!=\!1764\sim N\!=\!6400$. Relatively smaller number of MCS is done at non-transition points in each $N$.

\section{Results}\label{Results}
\begin{figure}[hbt]
\centering
\includegraphics[width=10cm]{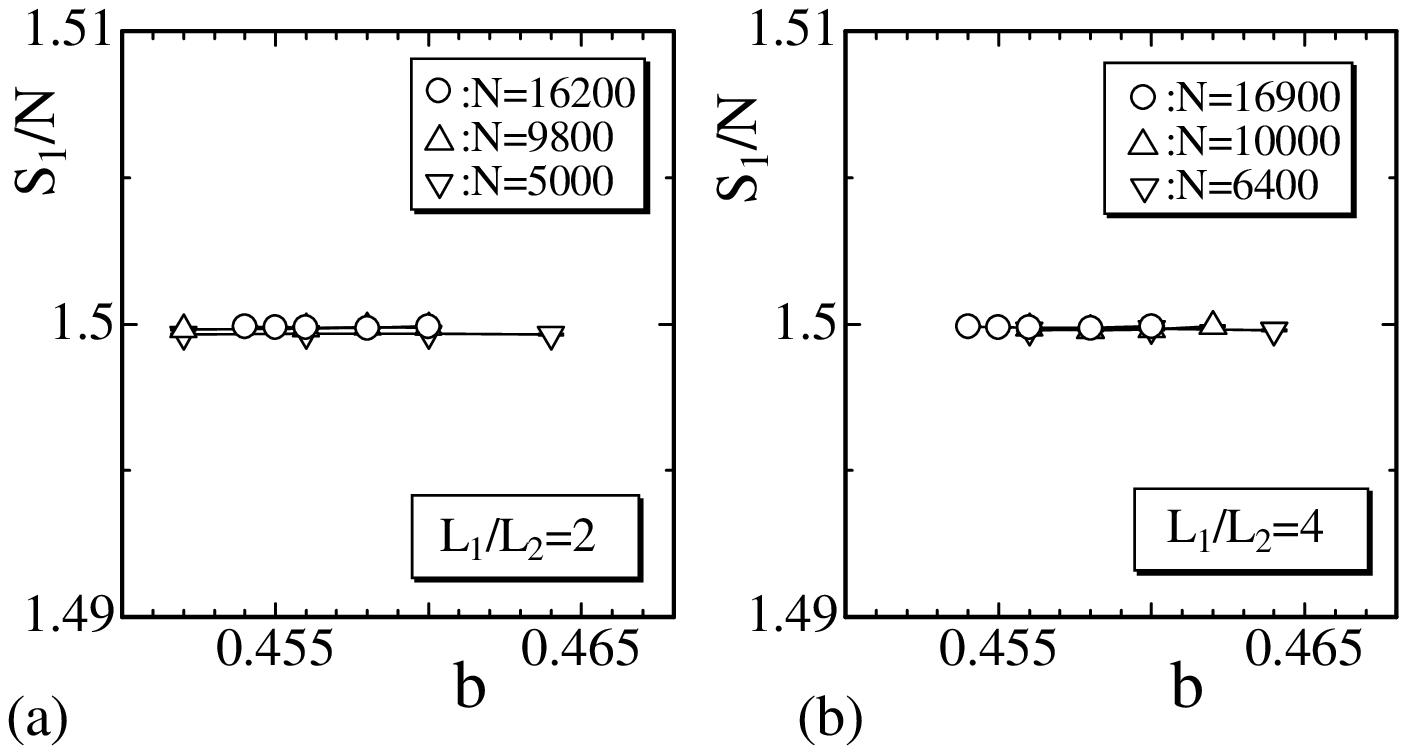}
\caption{$S_1/N$ vs. $b$ obtained on surfaces of (a) $N\!=\!5000\sim16200$, $L_1/L_2\!=\!2$, and (b)  $N\!=\!6400\sim 16900$, $L_1/L_2\!=\!4$. }
\label{fig-3}
\end{figure}
The relation $S_1/N\!=\!1.5$ is expected from the scale invariance of the partition function and can be used to check that the simulation was correctly performed. We also expect that this relation should not be influenced by whether the phase transition is of first order or not. Figures \ref{fig-3}(a) and \ref{fig-3}(b) are $S_1/N$ against $b$ obtained on surfaces of size $N\!=\!5000\sim 16200$, $L_1/L_2\!=\!2$ and $N\!=\!6400\sim 16900$, $L_1/L_2\!=\!4$. We find from these figures that the expected relation $S_1/N\!=\!1.5$ is satisfied.

\begin{figure}[hbt]
\centering
\includegraphics[width=10cm]{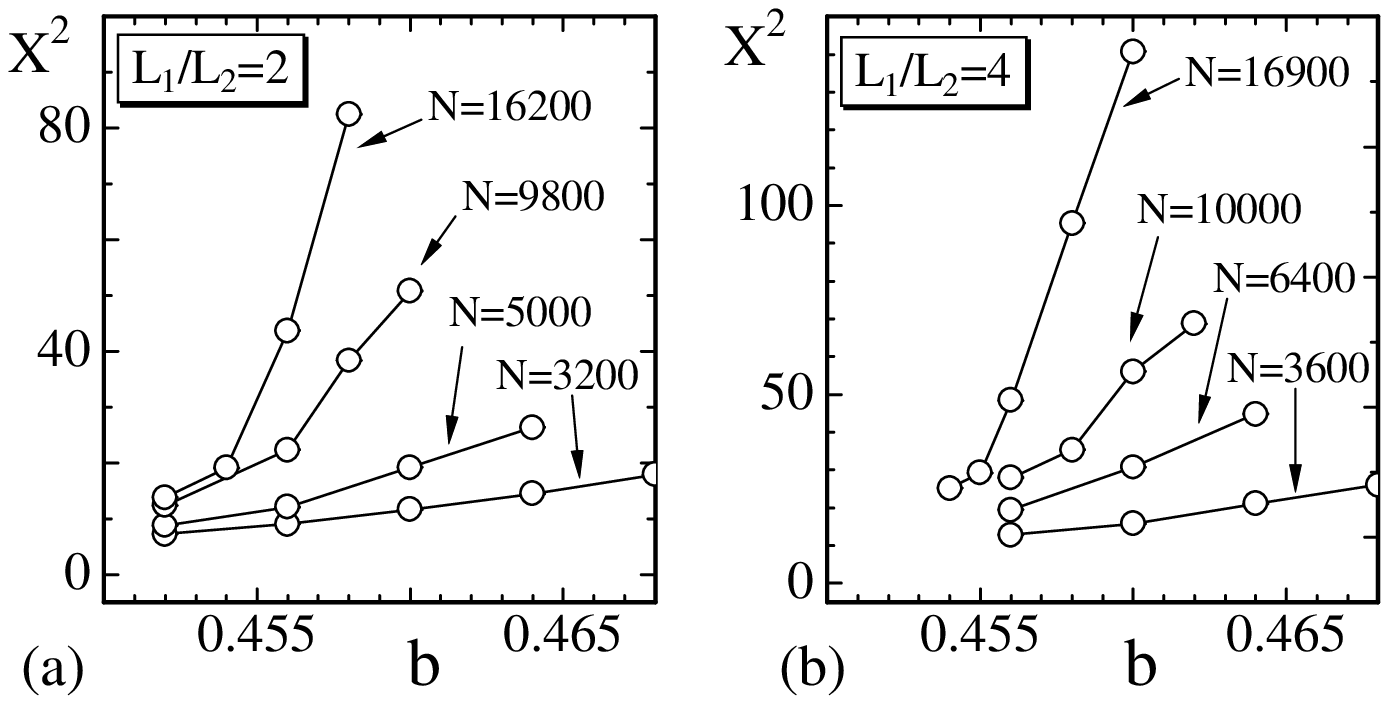}
\caption{$X^2$ vs. $b$ obtained on surfaces of (a) $N\!=\!5000\sim 16200$, $L_1/L_2\!=\!2$, and (b)  $N\!=\!6400\sim 16900$, $L_1/L_2\!=\!4$. }
\label{fig-4}
\end{figure}
The surfaces become smooth at $b\!\to\!\infty$ and crumpled at $b\!\to\!0$. Hence the size of the surfaces can be reflected in the mean square size $X^2$, which is defined by
\begin{equation}
\label{X2}
X^2={1\over N} \sum_i \left(X_i-\bar X\right)^2, \quad \bar X={1\over N} \sum_i X_i.
\end{equation}
In order to see the size of surfaces, we plot $X^2$ against $b$ in Fig. \ref{fig-4}(a). We see that $X^2$ grows rapidly at intermediate $b$, however it seems continuously change against $b$. We must note that the continuous behavior of $X^2$ does not always indicate a higher-order transition between the smooth phase and the crumpled phase. It is possible that $X^2$ changes continuously at the transition point, because the phase transition is characterized by the analytic structure of the partition function: the transition is of first-order (second-order) when ${\partial Z \over \partial b}$ is discontinuous (${\partial^2 Z \over \partial^2 b}$ is discontinuous).

\begin{figure}[hbt]
\centering
\includegraphics[width=10cm]{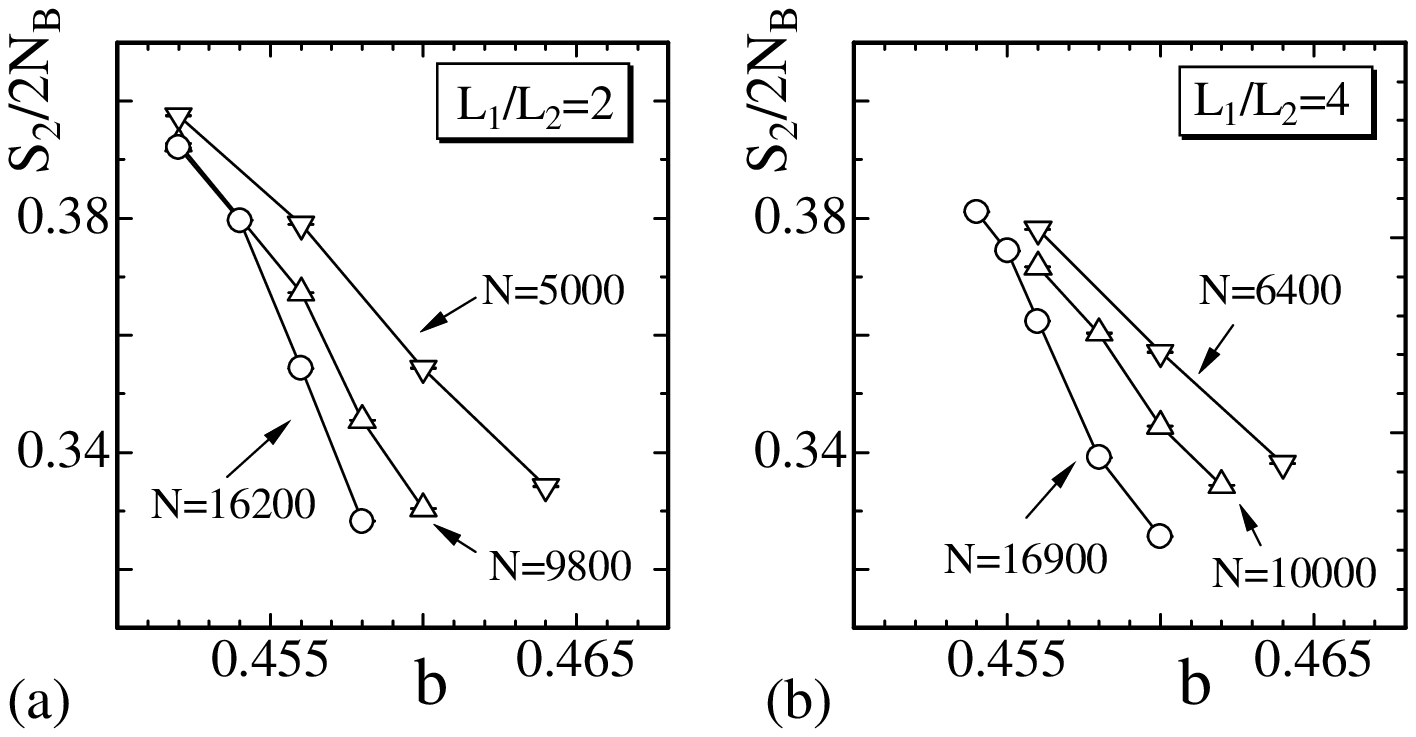}
\caption{The bending energy $S_2/2N_B$ vs. $b$ obtained on surfaces of (a) $N\!=\!5000\sim 16200$, $L_1/L_2\!=\!2$, and (b)  $N\!=\!6400\sim 16900$, $L_1/L_2\!=\!4$. $N_B$ is the total number of bonds.}
\label{fig-5}
\end{figure}
The bending energy $S_2/2N_B$ is plotted in Figs. \ref{fig-5}(a),(b). By dividing $S_2$ by $2N_B$, we obtain the bending energy per bond. In fact, the definition of $S_2$ in Eq. (\ref{S1S2})  includes the summation over bonds twice. $S_2/2N_B$ rapidly changes at $b$ where $X^2$ rapidly changes, however, discontinuous change is also hard to see in $S_2/2N_B$.

\begin{figure}[hbt]
\centering
\includegraphics[width=10cm]{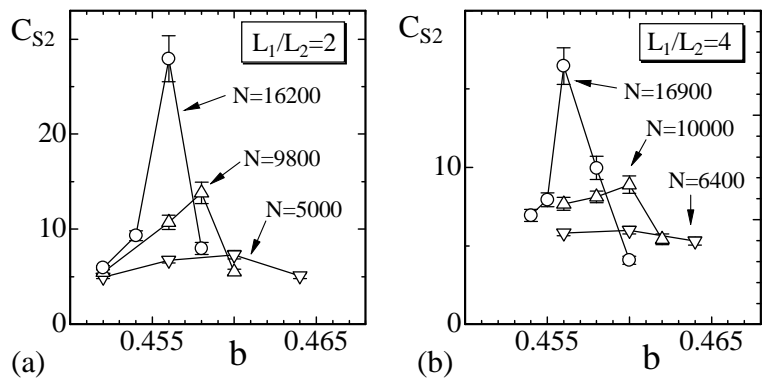}
\caption{The specific heat $C_{S_2}$ vs. $b$ obtained on surfaces of (a) $N\!=\!5000\sim 16200$, $L_1/L_2\!=\!2$, and (b)  $N\!=\!6400\sim 16900$, $L_1/L_2\!=\!4$. }
\label{fig-6}
\end{figure}
The specific heat $C_{S_2}$ is defined by
\begin{equation}
\label{Spec-Heat}
C_{S_2} = {b^2\over N} \langle \; \left( S_2 - \langle S_2 \rangle\right)^2 \; \rangle, 
\end{equation}
which reflects the phase transition if it anomalously behaves.   Figure \ref{fig-6}(a),(b)  shows $C_{S_2}$ against $b$, which were obtained on surfaces of $N\!=\!5000\sim 16200$, $L_1/L_2\!=\!2$, and  $N\!=\!6400\sim 16900$, $L_1/L_2\!=\!4$, respectively. Anomalous behaviors are obviously seen in $C_{S_2}$ in the figures. This indicates that the fluctuation of $S_2$ becomes very large at $b$ where $C_{S_2}$ has the peak at each $N$. Thus, we found a sign of a phase transition.

\begin{figure}[hbt]
\centering
\includegraphics[width=10cm]{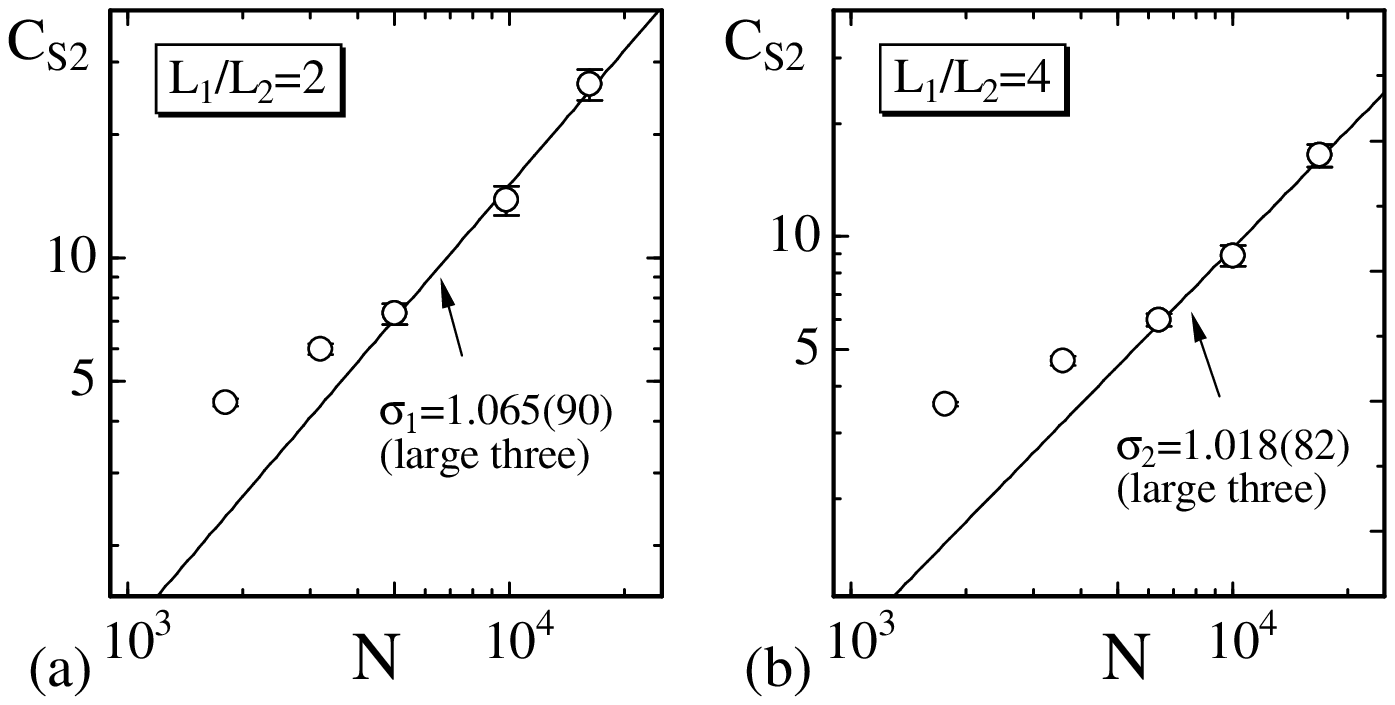}
\caption{Log-log plots of the peak value $C_{S_2}^{\rm max}$ against $N$, which were obtained on surfaces of (a) $N\!=\!1800\sim 16200$, $L_1/L_2\!=\!2$, and (b)  $N\!=\!1764\sim 16900$, $L_1/L_2\!=\!4$.  }
\label{fig-7}
\end{figure}
In order to see an order of the transition, we plot the peak value $C_{S_2}^{\rm max}$ against $N$ in Figs. \ref{fig-7}(a),(b) in a log-log scale. Figures include the simulation results obtained on surfaces of $N\!=\!1800$, $N\!=\!3200$ for $L_1/L_2\!=\!2$, and on $N\!=\!1764$, $N\!=\!3600$ for $L_1/L_2\!=\!4$. The straight lines in Figs. \ref{fig-7}(a),(b) were drawn by fitting the largest three $C_{S_2}^{\rm max}$ to 
\begin{equation}
\label{sigma-fitting}
C_{S_2}^{\rm max} \sim N^\sigma,
\end{equation}
where $\sigma$ is a critical exponent of the transition. Thus, we have
\begin{eqnarray}
\label{sigma-result}
\sigma_1=1.065\pm 0.090,\quad (L_1/L_2\!=\!2),\nonumber \\
\sigma_2=1.018\pm 0.082,\quad (L_1/L_2\!=\!4).
\end{eqnarray}
These results indicate that the phase transition is of first-order in both cases for $L_1/L_2\!=\!2$ and $L_1/L_2\!=\!4$. We also see that the model show a continuous transition on small surfaces of size $N\leq 5000\!\sim\! 7000$, since the exponent $\sigma$ is $\sigma \!<\! 1$ at $N\leq 5000\!\sim\! 7000$, which is obvious from the figures. Therefore, the torus is severly influenced by the size effect as expected, since the first-order transition can be seen in the same model on the sphere of $N\geq 1500\!\sim\!2500$ \cite{KOIB-PRE-2004-1}.

\begin{figure}[hbt]
\centering
\includegraphics[width=12cm]{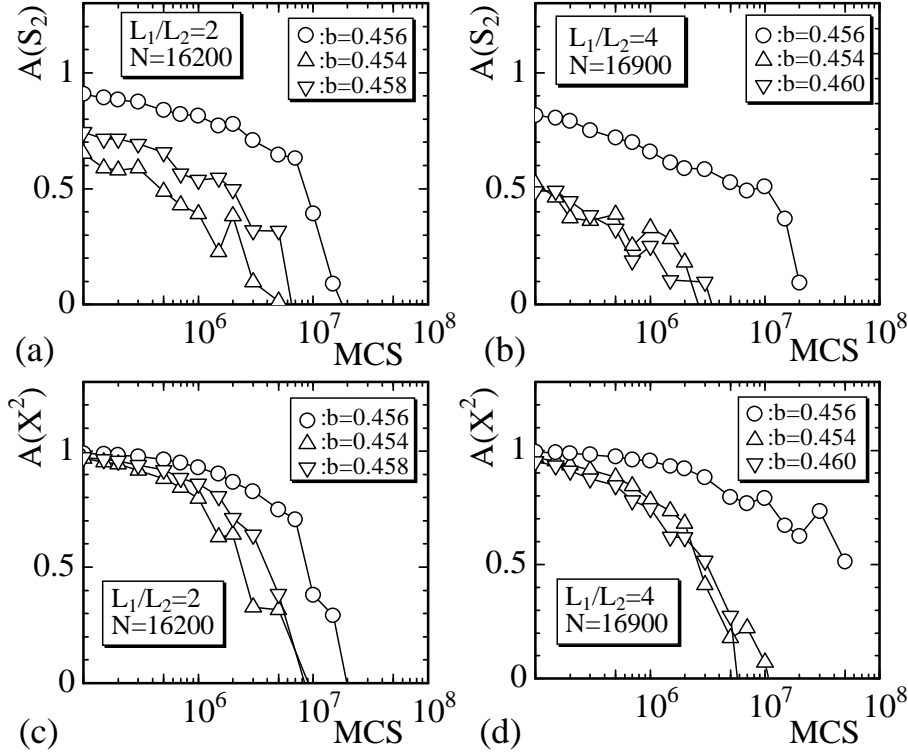}
\caption{Auto-correlation coefficient $A(S_2)$ obtained on surfaces of (a) $N\!=\!16200$, $L_1/L_2\!=\!2$, (b)  $N\!=\!16900$, $L_1/L_2\!=\!4$, and $A(X^2)$ obtained on surfaces of (c) $N\!=\!16200$, $L_1/L_2\!=\!2$, (d)  $N\!=\!16900$, $L_1/L_2\!=\!4$. }
\label{fig-8}
\end{figure}
Figures \ref{fig-8}(a),(b) are the autocorrelation coefficient $A(S_2)$ of $S_2$ defined by 
\begin{eqnarray}
A(S_2)= \frac{\sum_i S_2(\tau_{i}) S_2(\tau_{i+1})} 
   {  \left[\sum_i  S_2(\tau_i)\right]^2 },\\ \nonumber
 \tau_{i+1} = \tau_i + n \times 500, \quad n=1,2,\cdots.  
\end{eqnarray}
The autocorrelation coefficient $A(X^2)$ of $X^2$ were also plotted in Figs. \ref{fig-8}(c),(d). 
The phase transition can be reflected in the convergence speed of the MC simulations. The horizontal axes in the figure represent $500\!\times\! n\;(n\!=\!1,2,\cdots)$-MCS, which is a sampling-sweep between the samples $S_2(\tau_i)$ and $S_2(\tau_{i+1})$ . The coefficients $A(S_2)$ and $A(X^2)$ in Figs. \ref{fig-8}(a)--(d) were obtained on surfaces of size $N\!=\!16200$ for $L_1/L_2\!=\!2$ and $N\!=\!16900$ for $L_1/L_2\!=\!4$. The symbol ($\bigcirc $) represents data obtained at the transition point, and the symbols ($\bigtriangleup $) and ($\bigtriangledown $) represent those obtained in the crumpled phase and in the smooth phase, respectively.  

We clearly see from the figures that the convergence speed at the transition point ($\bigcirc $)  is about 10 times larger than those in the crumpled ($\bigtriangleup $) and the smooth ($\bigtriangledown $) phases both for $S_2$ and $X^2$. This is a phenomenon of the critical slowing down typical to the phase transition. The reason why the convergence speed at the transition point is so larger than in the crumpled and the smooth phases is that the phase space volume ($\subseteq {\bf R}^3$), where the vertices $X_i$ take their values, becomes larger at the transition point than that in the smooth and crumpled phases.

\begin{figure}[hbt]
\vspace{2cm}
\unitlength 0.1in
\begin{picture}( 10,10)(  0,0)
\put(15,40){\makebox(0,0){(a) Smooth surface}}%
\put(18.5,38){\makebox(0,0){$N\!=\!16200,\;L_1/L_2\!=\!2,$}}%
\put(14,36){\makebox(0,0){$b\!=\!0.456$}}%
\put(40,40){\makebox(0,0){(b) The surface section of (a)}}%
\put(41,-2){\makebox(0,0){(d) The surface section of (c)}}%
\put(16,-2){\makebox(0,0){(c) Crumpled surface}}%
\put(18.5,-4){\makebox(0,0){$N\!=\!16200,\;L_1/L_2\!=\!2,$}}%
\put(14,-6){\makebox(0,0){$b\!=\!0.456$}}%
\end{picture}%
\vspace{1.5cm}
\includegraphics[width=9.0cm]{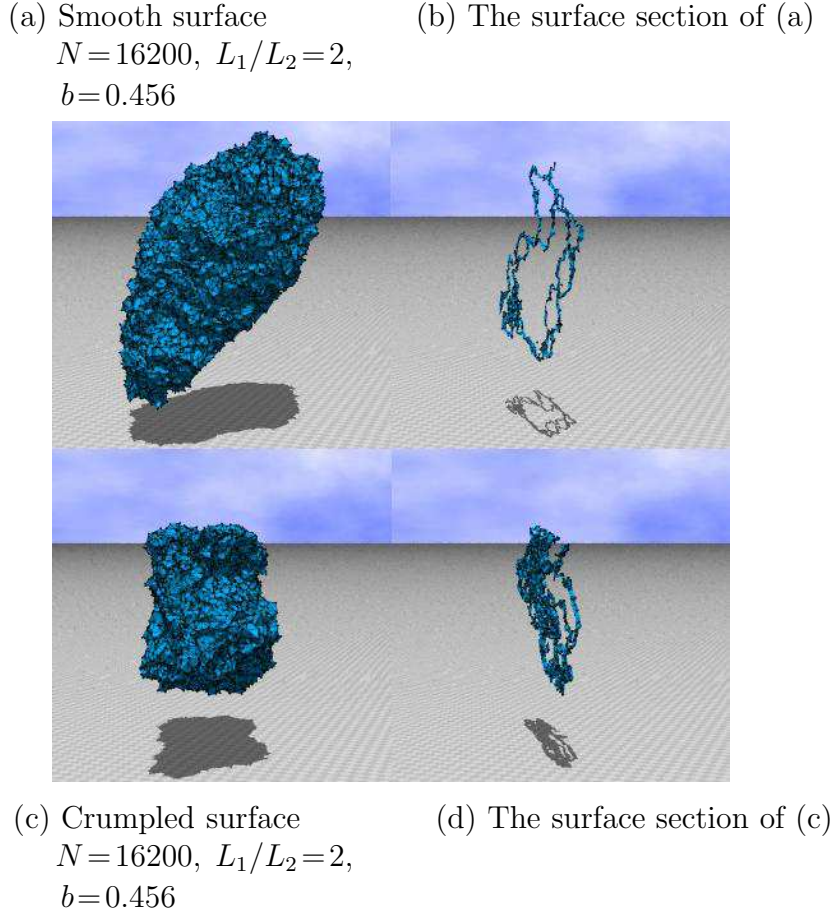}
\caption{ Snapshot of the surface of $N\!=\!16200$, $L_1/L_2\!=\!2$ obtained at the transition point $b\!=\!0.456$; (a) a smooth surface and (b) the surface section, (c) a crumpled surface and (d) the surface section. The mean square size of the smooth surface is $X^2\!=\!87$, and that of the crumpled one is  $X^2\!=\!32$.  }
\label{fig-9}
\end{figure}
Snapshots of surfaces are shown in Figs. \ref{fig-9}(a)--(d), which were obtained at the transition point $b\!=\!0.456$ on the $N\!=\!16200$, $L_1/L_2\!=\!2$ surface. The surface in Fig. \ref{fig-9}(a) is a smooth one, whose $X^2$ is about $X^2\!=\!87$. Figure \ref{fig-9}(b) is the section of the surface in Fig. \ref{fig-9}(a). The surface in Fig. \ref{fig-9}(c) is a crumpled one, whose $X^2$ is about $X^2\!=\!32$. Figure \ref{fig-9}(d) is the section of the surface in Fig. \ref{fig-9}(c). We can see from the surface section in Fig. \ref{fig-9}(b) that there is a lot of empty space inside the smooth surface in Fig. \ref{fig-9}(a). On the contrary, we can also see from Fig. \ref{fig-9}(d) that the space is obviously reduced in the crumpled surface in Fig. \ref{fig-9}(c). We note that $X^2$ seems continuously change at the transition point, in contrast to the model on a sphere \cite{KOIB-PRE-2004-1,KOIB-PRE-2005,KOIB-NPB-2006}, where $X^2$ has two distinct values at the transition point. The variation of $X^2$ against MCS, which was not depicted in a figure, indicates the continuous behavior of $X^2$ at the transition point.

\begin{figure}[hbt]
\vspace{2cm}
\unitlength 0.1in
\begin{picture}( 10,10)(  0,0)
\put(15,40){\makebox(0,0){(a) Smooth surface}}%
\put(18.5,38){\makebox(0,0){$N\!=\!16900,\;L_1/L_2\!=\!4,$}}%
\put(14,36){\makebox(0,0){$b\!=\!0.456$}}%
\put(40,40){\makebox(0,0){(b) The surface section of (a)}}%
\put(40.5,-2){\makebox(0,0){(d) The surface section of (c)}}%
\put(15.5,-2){\makebox(0,0){(c) Crumpled surface}}%
\put(18,-4){\makebox(0,0){$N\!=\!16900,\;L_1/L_2\!=\!4,$}}%
\put(13.5,-6){\makebox(0,0){$b\!=\!0.456$}}%
\end{picture}%
\vspace{1.5cm}
\includegraphics[width=9.0cm]{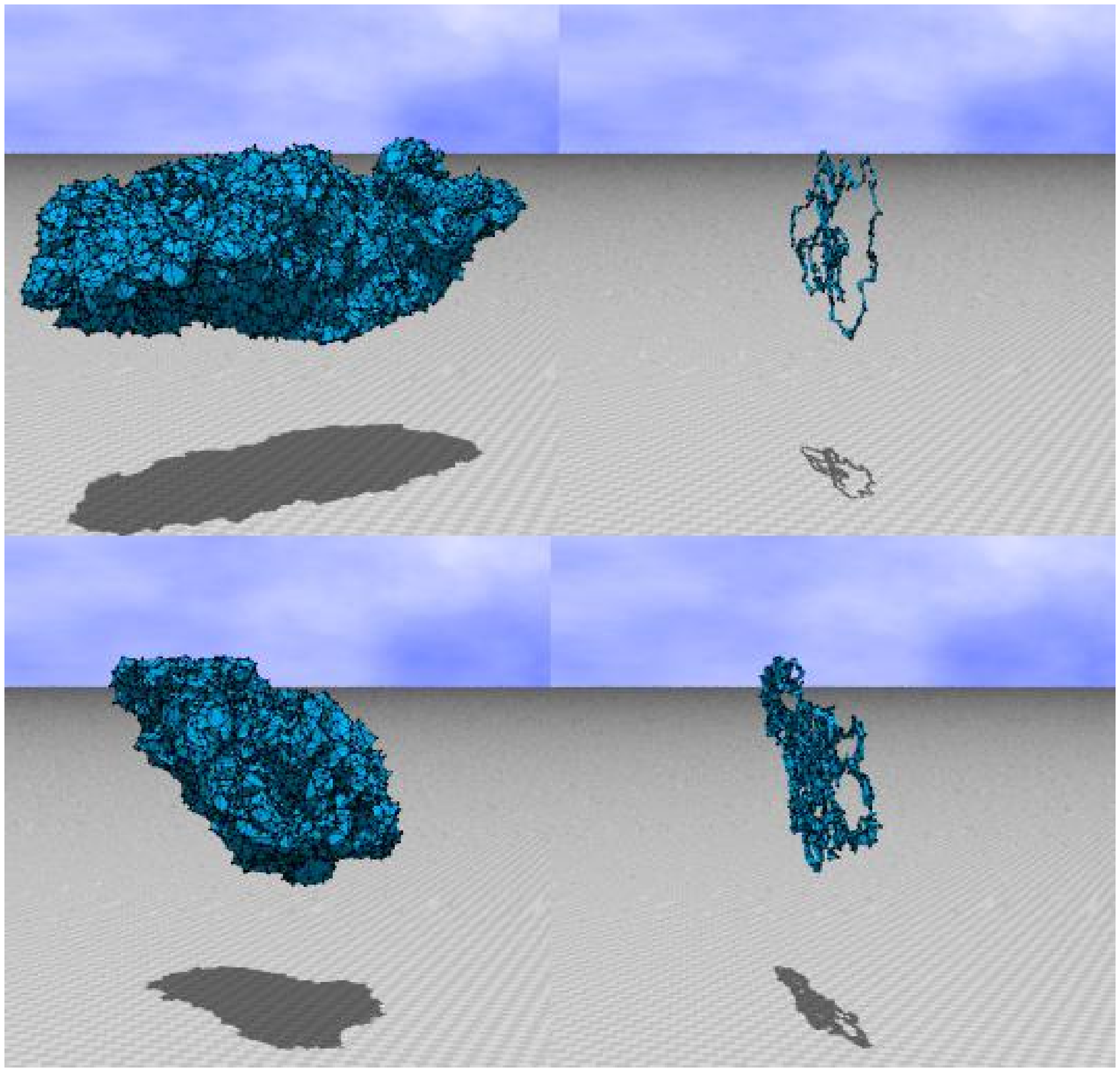}
\caption{Snapshot of the surface of $N\!=\!16900$, $L_1/L_2\!=\!4$ obtained at the transition point $b\!=\!0.456$; (a) a smooth surface and (b) the surface section, (c) a crumpled surface and (d) the surface section. The mean square size of the smooth surface is $X^2\!=\!98$, and that of the crumpled one is  $X^2\!=\!31$. }
\label{fig-10}
\end{figure}
Snapshots of surfaces are shown in Figs. \ref{fig-10}(a)--(d), which were obtained at the transition point $b\!=\!0.456$ on the $N\!=\!16900$, $L_1/L_2\!=\!4$ surface. The surface in Fig. \ref{fig-10}(a) is a smooth one, whose $X^2$ is about $X^2\!=\!98$. Figure \ref{fig-10}(b) is the section of the surface in Fig. \ref{fig-10}(a). The surface in Fig. \ref{fig-10}(c) is a crumpled one, whose $X^2$ is about $X^2\!=\!31$. Figure \ref{fig-10}(d) is the section of the surface in Fig. \ref{fig-10}(c). We can see from the surface section in Fig. \ref{fig-10}(b) that there is a lot of empty space inside the smooth surface in Fig. \ref{fig-10}(a). On the contrary, we can also see from Fig. \ref{fig-10}(d) that the space is obviously reduced in the crumpled surface in Fig. \ref{fig-10}(c). We note also that $X^2$ seems continuously change at the transition point, as in the case for $L_1/L_2\!=\!2$.

\section{Summary and conclusion}\label{Conclusions}
We have shown that a tethered surface model with an extrinsic curvature undergoes a first-order phase transition between the smooth phase and the crumpled phase on tori. The tori were constructed by connecting the sides of rectangular surfaces of size $L_1\times L_2$. The specific heat $C_{S_2}$ has the peak $C_{S_2}^{\rm max}$ at the transition point, and  $C_{S_2}^{\rm max}$ scales according to $C_{S_2}^{\rm max}\sim N^\sigma$, where $\sigma=1.065(90)$ for $L_1/L_2\!=\!2$ and $\sigma=1.018(82)$ for $L_1/L_2\!=\!2$. These results indicate the transition is of first-order on the basis of the finite-size scaling. We found an anomalous behavior of $C_{S_2}^{\rm max}$, which is consistent with the first-order transition on the surfaces of size $N\geq 5000\!\sim\! 7000$. The surface size  $N= 5000\!\sim\! 7000$ is very large compared with $N=1500\!\sim\!2500$ of the same model on a sphere \cite{KOIB-PRE-2004-1}. This implies that very large tori are necessary for the model with the standard bending energy to undergo the first-order transition, because the model with the standard bending energy on a sphere shows the first-order transition on the surfaces of $N\!\geq 7000$ \cite{KOIB-PRE-2005,KOIB-NPB-2006}.

It must be noted that a gap was not seen in $S_2$ at the transition point, and hence the histogram of $S_2$ has no double peak structure in contrast to the model on spheres. The reason of this seems due to the size effect. The model on the torus is severely affected by the size effect.  Nevertheless, we confirmed from the finite-size scaling that the model undergoes a fist-order transition as stated above.

Thus, the results of the extrinsic curvature model on tori in this Letter together with those previously obtained on the same class of model on spheres can lead us to conclude that the model undergoes a first-order transition between the smooth phase and the crumpled phase on {\it compact} surfaces. Moreover, the phase transition and the topology change is {\it compatible} in the extrinsic curvature model. 
The term {\it compatible} means that both of two phenomena lead to the same result without depending on which phenomenon firstly occurs, as discussed in \cite{KOIB-PLA-2006-1} for the intrinsic curvature model.

It was already reported that the intrinsic curvature model undergoes a first-order transition, which is independent of whether the surface is compact or noncompact. This fact and the result in this Letter imply that the tethered surface model has a first-order crumpling transition on compact surface, and the transition occurs independent of whether the curvature energy is an intrinsic one or an extrinsic one.  

This work is supported in part by a Grant-in-Aid for Scientific Research, No. 15560160.



\end{document}